\documentclass[pra,aps,showpacs,twocolumn]{revtex4}
\usepackage{graphicx}
\usepackage{latexsym}
\usepackage{amsmath}
\usepackage[dvipdfm]{hyperref}

\begin{document}

\title{Oblique discord}

\author{Jianwei Xu}
\email{xxujianwei@nwafu.edu.cn}
\affiliation{College of Science, Northwest A\&F University, Yangling, Shaanxi 712100, China}
\date{\today}

\begin{abstract}
Discord and entanglement characterize two kinds of quantum
correlations, and discord captures more correlation than entanglement in the
sense that even separable states may have nonzero discord. In this paper, we
propose a new kind of quantum correlation we call it oblique discord. A
zero-discord state corresponds to an orthonormal basis, while a
zero-oblique-discord state corresponds to a basis which is not necessarily
orthogonal. Under this definition, the set of zero-discord states is
properly contained inside the set of zero-oblique-discord states, and the
set of zero-oblique-discord states is properly contained inside the set of
separable states. We give a characterization of zero-oblique-discord states
via quantum operation, provide a geometric measure for oblique discord, and
raise a conjecture with it holds we can define an information-theoretic
measure for oblique discord. Also, we point out that, the definition of
oblique discord can be properly extended to some different versions just as
the case of quantum discord.

\end{abstract}


\maketitle

\bigskip

\section{Introduction}

Quantum correlation is one of the most striking features of quantum physics,
and leads to powerful applications in quantum information science
\cite{Horodecki2009,Modi2012}. Discord and
entanglement characterize two kinds of quantum correlations, manifest
complex structures and achieved fruitful results
\cite{Horodecki2009,Modi2012}. Discord captures more correlation than
entanglement in the sense that even separable states may have nonzero
discord, although for certain measures discord not necessarily is larger
than entanglement.

This paper asks the question: are there other kinds of correlation between
entanglement and discord. To this aim, we properly generalize the definition
of discord, we call the generalized version oblique discord. Under this
definition, the set of zero-discord states is properly contained inside the
set of zero-oblique-discord states, and the set of zero-oblique-discord
states is properly contained inside the set of separable states. Moreover,
we provide the information-theoretic measure and geometric measure for
oblique discord compared to the case of discord, and propose the definition
of global oblique discord compared to global discord.

This paper is organized as follows. In section 2, as preparations, we
review the definitions of entanglement, discord, geometric discord,
global discord and geometric global discord. In section 3, we provide the
definition of oblique discord, give a characterization of
zero-oblique-discord states via quantum operation, provide a geometric
measure for oblique discord. Also, we raise a conjecture, if it holds we can
define an information-theoretic measure for oblique discord. In section 4,
we point out that, the definition of oblique discord can be properly
extended to some different versions just as the case of quantum discord. In
section 5, we give a summary.

\section{Entanglement and discord}

Suppose the quantum systems A, B are described by the complex Hilbert spaces $H^{A}$
and $H^{B}$, $n_{A}=\dim H^{A}$ and $n_{B}=\dim H^{B}$ are finite. The
bipartite system $AB$ is then described by the Hilbert space $%
H^{AB}=H^{A}\otimes H^{B}$ with $\dim H^{AB}=n_{A}n_{B}.$ Let $I_{A},I_{B}$
be the identity operators of A and B, then the identity operator of AB is $%
I_{AB}=I_{A}\otimes I_{B}$. When we consider an $N$-partite system $%
A_{1}A_{2}...A_{N}$, we use $\{A_{i}\}_{i=1}^{N}$ to denote each subsystem
and their Hilbert spaces are $H^{A_{i}},$ the dimension $n_{A_{i}},$ the
identity $I_{A_{i}}$. We often omit the identity operator, for example we
write $\rho ^{A}\otimes I_{B}$ as $\rho ^{A}$ by omitting $I_{B}$, without
any confusion.

A quantum state $\rho ^{AB}$ is called a separable state if it can be
written in the form
\begin{eqnarray}
\rho ^{AB}=\sum_{i}p_{i}\rho _{i}^{A}\otimes \rho _{i}^{B},
\end{eqnarray}
where $\sum_{i}p_{i}=1,p_{i}\geq 0,\{\rho _{i}^{A}\}_{i}$ are states on $%
H^{A}$ and $\{\rho _{i}^{B}\}_{i}$ are states on $H^{B}$. $\rho ^{AB}$ is
called an entangled state or disentangled state if it is not separable. By
far many entanglement measures have been proposed \cite{Horodecki2009}.

A state $\rho ^{AB}$ is called a zero-discord state with respect to A if it
can be written in the form
\begin{eqnarray}
\rho ^{AB}=\sum_{\alpha }p_{\alpha }|\alpha \rangle \langle \alpha |\otimes
\rho _{\alpha }^{B},
\end{eqnarray}
where $\sum_{\alpha }p_{\alpha }=1,p_{\alpha }\geq 0,\{|\alpha \rangle
\}_{\alpha }$ is an orthonormal basis of $H^{A}$, $\{\rho _{\alpha
}^{B}\}_{\alpha }$ are states on $H^{B}$. $\rho ^{AB}$ is called a
discordant state if it is not a zero-discord state. The basic measure of
discord is the information-theoretic measure proposed by \cite{Ollivier2001, Henderson2001}, that is
\begin{eqnarray}
D^{A}(\rho ^{AB})=\inf_{\Pi _{A}}[I(\rho ^{AB})-I(\Pi _{A}\rho ^{AB})],
\end{eqnarray}
where, $I(\rho ^{AB})=S(\rho ^{A})+S(\rho ^{B})-S(\rho ^{AB})$ is mutual
information, $\rho ^{A}=tr_{B}\rho ^{AB}$ is reduced state, $S(\rho
^{AB})=-tr[\rho ^{AB}\log _{2}\rho ^{AB}]$ is entropy function, $\Pi _{A}$
denotes any projective measurement on A. It is shown \cite{Ollivier2001} that
\begin{eqnarray}
D^{A}(\rho ^{AB})\geq 0,  \ \ \ \ \ \ \ \ \ \ \ \ \ \ \ \ \ \ \ \ \ \ \ \ \ \ \ \ \ \ \    \\
D^{A}(\rho ^{AB})=0\Leftrightarrow \rho ^{AB}=\sum_{\alpha }p_{\alpha
}|\alpha \rangle \langle \alpha |\otimes \rho _{\alpha }^{B},
\end{eqnarray}
where $\sum_{\alpha }p_{\alpha }=1,p_{\alpha }\geq 0,\{|\alpha \rangle
\}_{\alpha }$ is an orthonormal basis of $H^{A}$, $\{\rho _{\alpha
}^{B}\}_{\alpha }$ are states on $H^{B}$. The intuitive meaning of $%
D^{A}(\rho ^{AB})$ is that it is the minimal loss of mutual information of $%
\rho ^{AB}$ over all projective measurement on A.

$D^{A}(\rho ^{AB})$ is difficult to get the analytical expressions except
for few special cases \cite{Luo2008}. Another
measure called geometric discord \cite{Dakic2010} is
defined as
\begin{eqnarray}
D_{G}^{A}(\rho ^{AB})=\inf \{tr[(\rho ^{AB}-\chi ^{AB})^{2}]:D^{A}(\chi
^{AB})=0\}.
\end{eqnarray}
It is obvious that
\begin{eqnarray}
D_{G}^{A}(\rho ^{AB})\geq 0,  \ \ \ \ \ \ \ \ \ \ \ \ \ \ \ \         \\
D_{G}^{A}(\rho ^{AB})\Leftrightarrow D^{A}(\rho ^{AB})=0.
\end{eqnarray}
For many cases, $D_{G}^{A}(\rho ^{AB})$ is easier to calculate than $%
D^{A}(\rho ^{AB})$ since $D_{G}^{A}(\rho ^{AB})$ avoids the complicated
entropy function. For instance, $D_{G}^{A}(\rho ^{AB})$ allows analytical
expressions for all $2\times d$ ($2\leq d<\infty $) states \cite{Dakic2010,Vinjanampathy2012}.

Discord with respect to A, $D^{A}(\rho ^{AB})$, can be extended to the
definition of global discord \cite{Rulli2011} that
(here we use the equivalent expression in \cite{Xu2013}
\begin{eqnarray}   
D(\rho ^{A_{1}A_{2}...A_{N}})=\inf_{\Pi _{A_{1}}\Pi _{A_{2}}...\Pi
_{A_{N}}}[I(\rho ^{A_{1}A_{2}...A_{N}})   \ \ \ \ \ \ \ \    \nonumber \\
-I(\Pi _{A_{1}}\Pi _{A_{2}}...\Pi
_{A_{N}}\rho ^{A_{1}A_{2}...A_{N}})],
\end{eqnarray}
where $I(\rho ^{A_{1}A_{2}...A_{N}})=\sum_{i=1}^{N}S(\rho ^{A_{i}})-S(\rho
^{A_{1}A_{2}...A_{N}})$ is the mutual information of $\rho
^{A_{1}A_{2}...A_{N}}$. $D(\rho ^{A_{1}A_{2}...A_{N}})$ has the property
\begin{eqnarray} 
D(\rho ^{A_{1}A_{2}...A_{N}})=0      \ \ \  \ \ \ \ \ \ \ \  \ \ \ \ \ \ \ \  \ \ \ \ \ \ \ \ \ \  \ \ \ \ \ \ \ \     \nonumber  \\
\Leftrightarrow \rho
^{A_{1}A_{2}...A_{N}}=\sum_{i_{1}=1}^{n_{1}}\sum_{i_{2}=1}^{n_{2}}...%
\sum_{i_{N}=1}^{n_{N}}p_{i_{1}i_{2}...i_{N}}|\alpha _{i_{1}}\rangle \langle
\alpha _{i_{1}}|  \nonumber \\
\otimes |\alpha _{i_{1}}\rangle \langle \alpha
_{i_{1}}|\otimes ...\otimes |\alpha _{i_{N}}\rangle \langle \alpha _{i_{N}}|,
\end{eqnarray}
where $p_{i_{1}i_{2}...i_{N}}\geq
0,\sum_{i_{1}i_{2}...i_{N}}p_{i_{1}i_{2}...i_{N}}=1,\{|\alpha
_{i_{j}}\rangle \}_{i_{j}=1}^{n_{j}}$ is an orthonormal basis of $H^{A_{j}}.$
For certain special states, $D(\rho ^{A_{1}A_{2}...A_{N}})$ possess analytical
expressions \cite{Rulli2011,Xu2013}.

Geometric discord with respect to A, $D_{G}^{A}(\rho ^{AB})$, can be
extended to the definition of geometric global discord \cite{Xu2012} that
\begin{eqnarray} 
D_{G}(\rho ^{A_{1}A_{2}...A_{N}})=\inf \{tr[(\rho
^{A_{1}A_{2}...A_{N}}-\chi ^{A_{1}A_{2}...A_{N}})^{2}]:  \nonumber \\
D(\rho^{A_{1}A_{2}...A_{N}})=0\}.   \ \ \ \ \
\end{eqnarray}
For certain special states, $D_{G}(\rho ^{A_{1}A_{2}...A_{N}})$ possess
analytical expressions \cite{Xu2012}.

The definitions of different kinds of discord above are all associated with
the projective measurements. Projective measurements are a kind of most
important quantum operations, but not all quantum operations. There are many
important problems, such as the optimal scheme to distinguish a set of
quantum states, involve other quantum operations, rather than a projective
measurement. A quantum operation is a map which maps a quantum state into
another quantum state \cite{Nielsen2000}. The familiar examples are projective measurement,
general measurement, amplitude damping and phase damping of qubit, etc. In
this paper, we relax the constraint of projective measurement, and seek a
more general definition other than quantum discord.

\section{Oblique discord and its measures}

\textbf{Definition 1}. We call the bipartite state $\rho
^{AB}$ a zero-oblique-discord state with respect to A, if $\rho ^{AB}$ can
be written in the form
\begin{eqnarray}
\rho ^{AB}=\sum_{i=1}^{n_{A}}p_{i}|i\rangle \langle i|\otimes \rho
_{i}^{B},
\end{eqnarray}
where $\sum_{i}p_{i}=1,p_{i}\geq 0,\{|i\rangle \}_{i=1}^{n_{A}}$ is a
normalized basis of $H^{A}$, $\{\rho _{i}^{B}\}_{i}$ are states on $H^{B}$.
Notice that $\{|i\rangle \}_{i=1}^{n_{A}}$ is not necessarily orthogonal.

Under this definition, combining Eqs.(1,2,12), we see that, the set of zero-discord states is
properly contained inside the set of zero-oblique-discord, and the set of
zero-oblique-discord states is properly contained inside the set of
separable states, see Fig.1.
\begin{figure} [!h]
\includegraphics[width=8cm,trim=100 550 0 50]{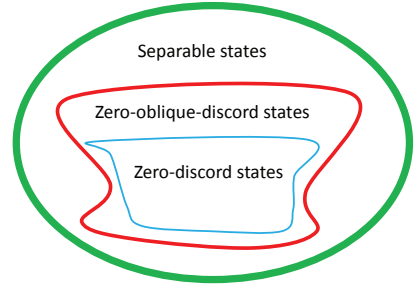}
\caption{Inclusion relations between the sets of  separable \\ states, zero-oblique-discord states and zero-discord states.}
\end{figure}

We give a characterization of zero-oblique-discord states via quantum
operations.
Suppose $\{|i\rangle \}_{i=1}^{n_{A}}$ is a normalized basis of $H^{A}$
which not necessarily orthogonal to each other. There exists an unique basis
$\{|\widetilde{i}\rangle \}_{i=1}^{n}$ of $H^{A}$ such that $\langle i|%
\widetilde{j}\rangle =\delta _{ij}$, note that $\{|\widetilde{i}\rangle
\}_{i=1}^{n}$ not necessarily orthogonal and not necessarily normalized. $\{|%
\widetilde{i}\rangle \}_{i=1}^{n_{A}}$ is called the dual basis of $%
\{|i\rangle \}_{i=1}^{n_{A}}$. We define the quantum operation $\Phi
_{A}=\{|i\rangle \langle \widetilde{i}|\}_{i=1}^{n_{A}}$ which operates the
bipartite state $\rho ^{AB}$ as
\begin{eqnarray}
\Phi _{A}\rho ^{AB}=\frac{\sum_{i=1}^{n_{A}}|i\rangle \langle \widetilde{i}%
|\rho ^{AB}|\widetilde{i}\rangle \langle i|}{tr[\sum_{i=1}^{n_{A}}\langle
\widetilde{i}|\rho ^{AB}|\widetilde{i}\rangle ]}. \label{eq.13}
\end{eqnarray}
With this definition, we have Theorem 1 below.

\textbf{Theorem 1.} A bipartite state $\rho ^{AB}$ is a
zero-oblique-discord state with respect to A, iff there exists an operation $%
\Phi _{A}=\{|i\rangle \langle \widetilde{i}|\}_{i=1}^{n_{A}}$ defined as in
Eq.(\ref{eq.13}) such that
\begin{eqnarray}
\Phi _{A}\rho ^{AB}=\rho ^{AB}.
\end{eqnarray}
\textbf{Proof.} Suppose there exists an operation $\Phi _{A}=\{|i\rangle
\langle \widetilde{i}|\}_{i=1}^{n_{A}}$ such that $\Phi _{A}\rho ^{AB}=\rho
^{AB}$, we expand $\rho ^{AB}$ as
\begin{eqnarray}
\rho ^{AB}=\sum_{jk=1}^{n_{A}}\sum_{\lambda \mu =1}^{n_{B}}\rho
_{jk,\lambda \mu }|j\rangle \langle k|\otimes |\lambda \rangle \langle \mu
|,
\end{eqnarray}
where $\{|j\rangle \}_{j=1}^{n_{A}}=\{|k\rangle \}_{k=1}^{n_{A}}=\{|i\rangle
\}_{i=1}^{n_{A}}$, $\{|\lambda \rangle \}_{\lambda =1}^{n_{B}}=\{|\mu
\rangle \}_{\mu =1}^{n_{B}}$ is an orthonormal basis of $H^{B}$, $\rho
_{jk,\lambda \mu }=\langle \widetilde{j}\lambda |\rho ^{AB}|\widetilde{k}\mu
\rangle $. Eq.(14) then reads
\begin{eqnarray}
\rho ^{AB}=\frac{\sum_{i=1}^{n_{A}}\sum_{\lambda \mu =1}^{n_{B}}\rho
_{ii,\lambda \mu }|i\rangle \langle i|\otimes |\lambda \rangle \langle \mu |%
}{\sum_{i=1}^{n_{A}}\sum_{\lambda =1}^{n_{B}}\rho _{ii,\lambda \lambda }},
\end{eqnarray}
it is of the form in Eq.(12).

Conversely, suppose $\rho ^{AB}$ can be expressed by Eq.(12), then $\Phi
_{A}=\{|i\rangle \langle \widetilde{i}|\}_{i=1}^{n_{A}}$ fulfils $\Phi
_{A}\rho ^{AB}=\rho ^{AB}.$   \ \ \ \ \ \ \ \ \ \ \ \      $\Box$

Compared to the geometric measure of discord in Eq.(6), we propose the
definition of geometric oblique discord as follows.

\textbf{Definition 2.} We define the geometric oblique discord of the bipartite state $%
\rho ^{AB}$ with respect to $A$ as
\begin{eqnarray} 
D_{GO}^{A}(\rho ^{AB})=\inf_{\chi ^{AB}}\{d(\rho ^{AB},\chi ^{AB}): \ \ \ \ \ \ \ \ \ \ \ \ \ \ \ \ \ \ \ \ \ \      \nonumber \\ \ \ \ \ \ \ \ \ \ \ \ \ \ \ \ \
\chi^{AB} \text{ is a zero-oblique-discord state}\},
\end{eqnarray}
where, $inf$ runs over all zero-oblique-discord states $\chi ^{AB}$, $d$ is
a distance, for example,
\begin{eqnarray}
d(\rho ^{AB},\chi ^{AB})=tr[(\rho ^{AB}-\chi ^{AB})^{2}].
\end{eqnarray}

\textbf{Definition 3.}
In the same spirit of Ref. \cite{Luo2010}, we can
also define another geometric oblique discord as
\begin{eqnarray}
D_{GO1}^{A}(\rho ^{AB})=\inf_{\Phi _{A}}\{d[\rho ^{AB},\Phi _{A}(\rho
^{AB})]:    \ \ \ \ \ \ \ \ \ \ \ \ \      \nonumber \\ \ \ \ \ \ \ 
\Phi _{A} \text{ is defined in Eq.(\ref{eq.13})} \},
\end{eqnarray}
once more for example,
\begin{eqnarray}
d[\rho ^{AB},\Phi _{A}(\rho ^{AB})]=tr\{[\rho ^{AB}-\Phi _{A}(\rho
^{AB})]^{2}\}.
\end{eqnarray}

\textbf{Definition 4.}
Compared to the information-theoretic measure of discord in Eq.(3), it is
very desirable to define an information-theoretic measure of oblique discord
as
\begin{eqnarray}
D_{O}^{A}(\rho ^{AB})=\inf_{\Phi _{A}}[I(\rho )-I(\Phi _{A}\rho )],
\end{eqnarray}
where $I(\rho )=S(\rho ^{A})+S(\rho ^{B})-S(\rho )$ is the mutual
information.

However, we do not know whether $D_{O}^{A}(\rho ^{AB})$ defined
above is always nonnegative. Note that $D_{O}^{A}(\rho ^{AB})\geq 0$ iff $%
I(\rho ^{AB})\geq I(\Phi _{A}\rho ^{AB})$ for any $\Phi _{A}$. We raise the
conjecture below.

\textbf{Conjecture:}
\begin{eqnarray}
I(\rho ^{AB})\geq I(\Phi _{A}\rho ^{AB}) \text{ for any } \Phi _{A} \text{ and any } \rho ^{AB},
\end{eqnarray}
 where $I(\rho ^{AB})$ is the mutual information, $\Phi
_{A}$ is defined in Eq.(\ref{eq.13}).

\section{Extend oblique discord in some ways}

As in the case of discord, we can extend the definition of oblique discord
in many ways.

\textbf{Definition 5.} An $N$-partite state $\rho $ is said to be of zero global
oblique discord, if it can be written in the form
\begin{eqnarray}
\rho=\sum_{i_{1}=1}^{n_{A_{1}}}\sum_{i_{2}=1}^{n_{A_{2}}}...%
\sum_{i_{N}=1}^{n_{A_{N}}}p_{i_{1}i_{2}...i_{N}}|i_{1}\rangle \langle
i_{1}|   \ \ \ \ \ \ \ \ \ \ \ \ \      \nonumber \\ \ \ \ \ \ \
\otimes |i_{2}\rangle \langle i_{2}|\otimes ...\otimes |i_{N}\rangle
\langle i_{N}|,
\end{eqnarray}
where $\sum_{i_{1}=1}^{n_{A_{1}}}\sum_{i_{2}=1}^{n_{A_{2}}}...%
\sum_{i_{N}=1}^{n_{A_{N}}}p_{i_{1}i_{2}...i_{N}}=1,p_{i_{1}i_{2}...i_{N}}%
\geq 0,\{|i_{j}\rangle \}_{i_{j}=1}^{n_{A_{j}}}$ is a normalized basis of $%
H^{A_{j}}.$ Notice that $\{|i_{j}\rangle \}_{i_{j}=1}^{n_{A_{j}}}$ is not
necessarily orthogonal.

\textbf{Theorem 2.} An $N$-partite state $\rho $ is of zero global oblique discord
iff there exists an operation $\{\Phi _{A_{j}}\}_{j=1}^{N}=\{\{|i_{j}\rangle
\langle \widetilde{i_{j}}|\}_{i_{j}=1}^{n_{A_{j}}}\}_{j=1}^{N}$ such that
\begin{eqnarray}
\Phi _{A_{1}A_{2}...A_{N}}\rho =\Phi _{A_{1}}...\Phi _{A_{N-1}}\Phi
_{A_{N}}\rho =\rho.
\end{eqnarray}

It can be directly checked that
\begin{eqnarray}
\Phi _{A_{1}}(\Phi _{A_{2}}\rho )=\Phi _{A_{2}}(\Phi _{A_{1}}\rho ),
\end{eqnarray}
hence $\Phi _{A_{1}A_{2}...A_{N}}\rho =\Phi _{A_{1}}...\Phi _{A_{N-1}}\Phi
_{A_{N}}\rho $ above can be defined without any ambiguity.

\textbf{Definition 6.} We define the geometric global oblique discord of $N$-partite
state $\rho $ as
\begin{eqnarray}
D_{GO}(\rho )=\inf_{\Phi _{A_{1}A_{2}...A_{N}}}d[\rho,\Phi
_{A_{1}A_{2}...A_{N}}(\rho )],
\end{eqnarray}
where, d is a distance as in Eq.(7).

\textbf{Definition 7.} We define the global oblique discord of an N-partite state $\rho
$ as
\begin{eqnarray}
D_{O}(\rho )=\inf_{\Phi _{A_{1}A_{2}...A_{N}}}[I(\rho )-I(\Phi
_{A_{1}A_{2}...A_{N}}\rho )],
\end{eqnarray}
where $I(\rho )=\sum_{j=1}^{N}S(\rho ^{A_{j}})-S(\rho )$ is the mutual
information.  Similar to the case of Eq.(21), $D_{O}(\rho )\geq 0$ requires
Eq.(22) holds.

\section{Summary and discussion}

The definition of quantum discord corresponds to orthogonal basis, in this
paper, we relaxed the constraint of orthogonality, and proposed the
definition of oblique discord. Oblique discord characterizes a new kind of
quantum correlation between entanglement and discord.

There left many open questions for future investigations. Firstly, are there
physical effects which can be revealed by oblique discord? Secondly,
conjecture in Eq.(22) is true or false? Thirdly, how to calculate the
different measures of oblique discord analytically or efficiently
numerically, especially for $n$-qubit states.

This work was supported by the National Natural Science Foundation of
China (Grant No.11347213) and the Chinese Universities Scientific Fund (Grant No.2014YB029). The author thanks
Kai-Liang Lin and Lin Zhang for helpful discussions.

\end{document}